\documentclass[a4paper]{report}
\usepackage[utf8]{inputenc}
\usepackage[T1]{fontenc}
\usepackage{graphicx}
\usepackage{RJournal}
\usepackage{booktabs}

\begin{document}

%% do not edit, for illustration only
\sectionhead{Contributed research article}
\volume{XX}
\volnumber{YY}
\year{20ZZ}
\month{AAAA}

\begin{article}
% !TeX root = RJwrapper.tex
\title{\texttt{tsdataleaks}: An R Package to Detect Potential Data Leaks in Forecasting Competitions }
\author{by Thiyanga S. Talagala, Department of Statistics, Faculty of Applied Sciences, \newline University of Sri Jayewardenepura \newline ttalagala@sjp.ac.lk}

\maketitle

\abstract{%
Forecasting competitions are of increasing importance as a means to learn best practices and gain knowledge. Data leakage is one of the most common issues that can often be found in competitions. Data leaks can happen when the training data contains information about the test data. There are a variety of different ways that data leaks can occur with time series data. For example: i) randomly chosen blocks of time series are concatenated to form a new time series; ii) scale-shifts; iii) repeating patterns in time series; iv) white noise is added to the original time series to form a new time series, etc. This work introduces a novel tool to detect these data leaks. The tsdataleaks package provides a simple and computationally efficient algorithm to exploit data leaks in time series data. This paper demonstrates the package design and its power to detect data leakages with an application to forecasting competition data.
}

\hypertarget{Statement of Need}{%
\section{Statement of Need}\label{introduction}}

Time series forecasting competitions have played a significant role in the advancement of forecasting practices. Typically, in forecasting competitions, a collection of time series is given to the participants, and then the participants submit the forecasts for the required test period for each time series. During the competition period, only the training set for each time series is given to the public, and the test set is kept private from the public. Finally, competition organizers evaluate the forecast accuracy by comparing each participants submitted forecast values against the actual test period values. Participating in forecasting competitions not only aids in the identification of novel methods and facilitates their performance comparison against existing state-of-the-art forecasting techniques, as highlighted by \citealt{hyndman2020brief}, but also provides empirical evidence crucial for enhancing forecasting performance and advancing the theory and practice of forecasting \citep{makridakis2022m5}.

\begin{figure}[htbp]
\centering
\includegraphics[height=0.6\textwidth]{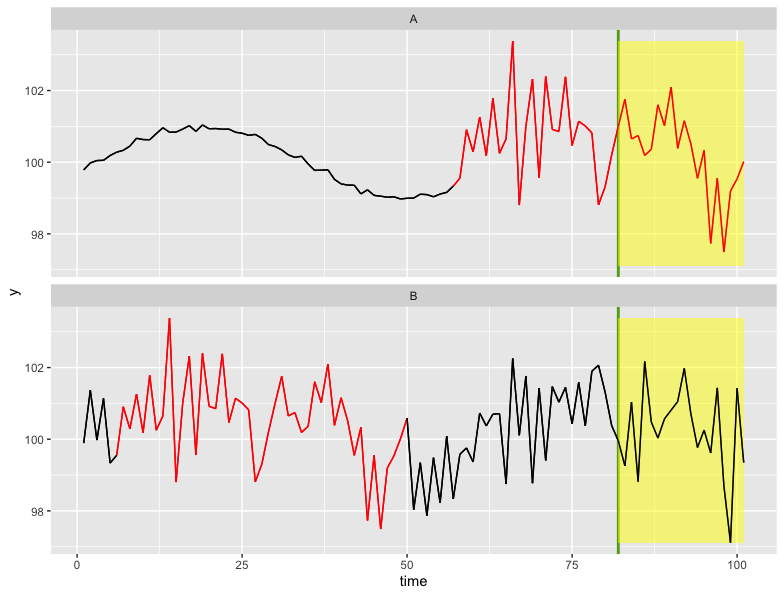}
\caption{An example of a time series data leak. "A" and "B"" are two time series. The green vertical line and yellow background separate the training and test parts of the series. A training segment of series B (red colour segment) is the source of the latter segment of the training set and test set of the A series.}
\label{fig:fig1}
\end{figure}

 \newpage

Data leakage occurs when the training period of the time series includes test period data before officially releasing the test period of the time series.  This idea is illustrated in \autoref{fig:fig1}. A and B are two time series. The latter segment of the training set and the subsequent test set within the (A) series are the same as the red segment highlighted in the training segment inherent to series (B). This type of data leak could occur when randomly chosen blocks of time series are concatenated to form a new time series.

Competitions with data leaks will not be able to reach their original purpose. By exploiting data leakage, competitors can obtain a top rank in the leader board. Such models look highly accurate within the competition environment but become inaccurate when applied to a data set outside the competition environment. Hence, there is an increasing need to examine the potential data leaks in time series before the release of data to the public. The tsdataleaks package is designed to identify data leaks in time series.

\hypertarget{State of the Field in R}{%
\section{State of the Field in R}\label{field}}

As of the latest information available on the Comprehensive R Archive Network (CRAN) Task View: Time Series Analysis \citep{ctv}, there is no package available for detecting data leakages.

\hypertarget{Algorithm}{%
\section{Algorithm}\label{Algorithm}}

The algorithm operates as follows: the algorithm begins by selecting the final segment of the training section in each time series within the collection. Subsequently, those segments iterates through all of the time series one time step at a time and calculate Pearson's correlation coefficient. Hence, the inputs to the algorithm are: i) the time series collection; ii) segment length; and iii) the cut-off value for the correlation coefficient. A data leak is indicated if the Pearson's correlation coefficient's absolute value exceeds the cutoff value. The algorithm returns the starting and end indexes of the segments that match each time series'  last segment corresponds to the training part.

\begin{table}[htbp]
\centering
\caption{Algorithm: Time Series Matching}
\label{alg:time_series_matching}
\begin{tabular}{p{15cm}}
\hline
\textbf{Input:} \\
1. \textit{lstx}: A collection of time series as a list in R. \\
2. \textit{h}: Length of the segment to be considered. \\
3. \textit{cutoff}: Cut-off value for the absolute value of the Pearson's correlation coefficient. \\
\textbf{Output:} \\
A list containing starting and ending indices of segments that match each time series' last segment of the training set. \\
\textbf{Steps:} \\
1. Extract the last segment of the training part with length \textit{h}. \\
2. Loop through the time series with one time step, considering each segment: \\
\quad - Calculate the Pearson's correlation coefficient between the current segment of a time series and the last portion of the training segment that was retrieved. \\
\quad - If the correlation coefficient is above the \textit{cutoff}: \\
\quad - Return the matching segments list with the starting and ending indices of the matching segments. \\
3. Return the matching segments list as the output. \\
\hline
\end{tabular}
\end{table}

\begin{figure}[htbp]
\centering
\includegraphics[width=1\textwidth]{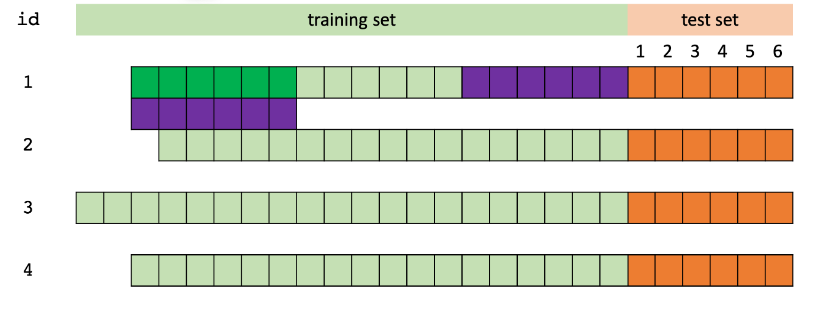}
\caption{Visualization of the first iteration of the algorithm. The last segment of the training part of the first series is colored purple. As the first step of the algorithm, it computes Pearson's correlation coefficient between the observations 1-6 (dark green section) and the purple segment.}
 \label{fig:fig21}
\end{figure}

\autoref{fig:fig21} illustrates the first iteration of the algorithm. The correlation between the purple segment and observations 1–6 (dark green section) of the first time series is measured at the first iteration.
 
\begin{figure}[htbp]
\centering
\includegraphics[width=1\textwidth]{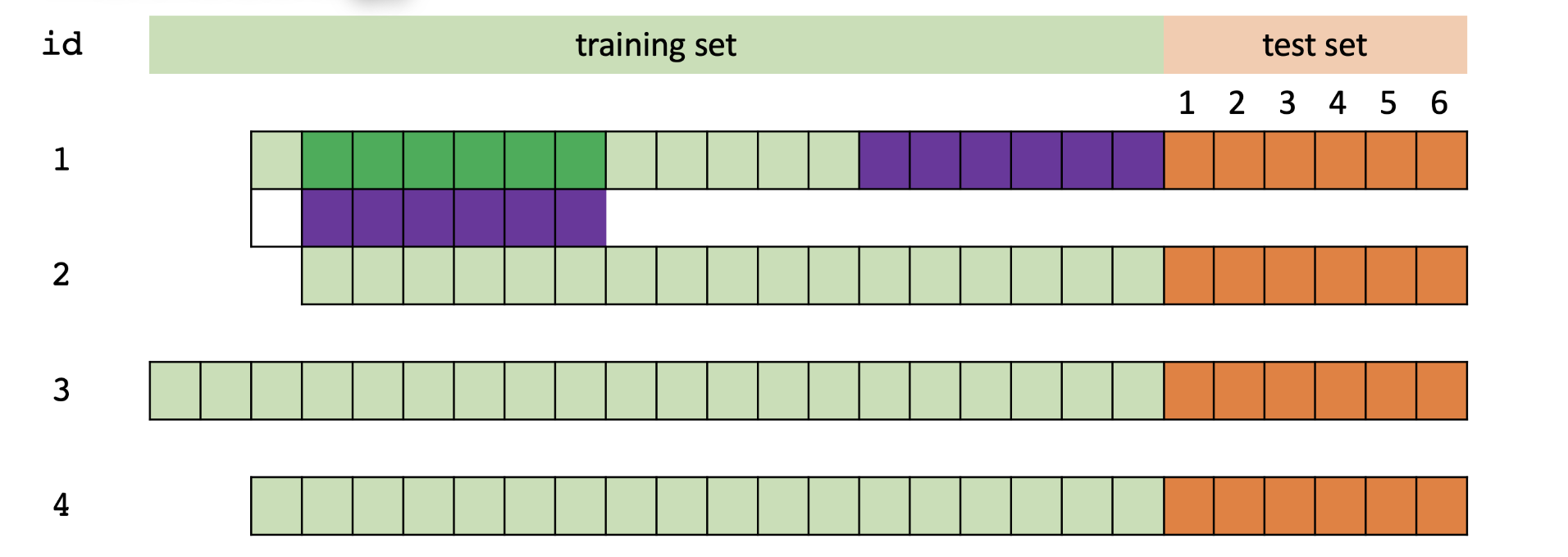}
\caption{Visualization of the second iteration of the algorithm. The last segment of the training part of the first series is colored purple. As the second step of the algorithm, it computes Pearson's correlation coefficient between the observations 2-7 (dark green section) and the purple segment.}
 \label{fig:fig22}
\end{figure}

\begin{figure}[htbp]
\centering
\includegraphics[width=1\textwidth]{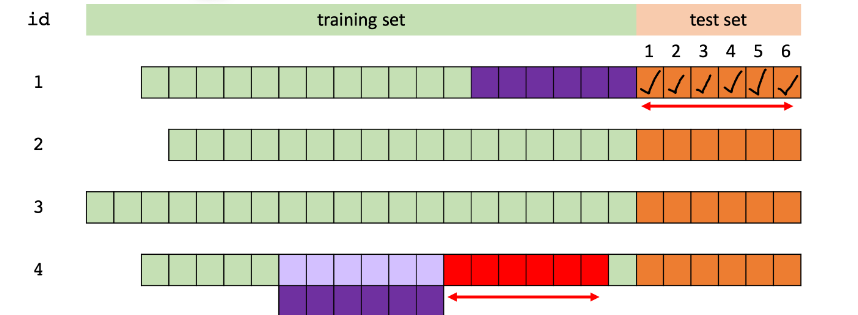}
\caption{Intermediate step of the algorithm: identification of potential data leaks. The light purple section of the fourth series perfectly correlates with the last segment of the first series. Hence, the red section of the fourth series could be the test part of the first series.}
 \label{fig:fig23}
\end{figure}

\autoref{fig:fig22} visualize the second iteration of the algorithm. The correlation between the purple segment and observations 2–7 (dark green section) of the first time series is measured at the second iteration. \autoref{fig:fig23} illustrates an intermediate step of the algorithm.

\newpage

\hypertarget{Usge}{%
\section{Usage}\label{Usage}}

\subsection{Installation}\label{Installation}

The package tsdataleaks is available on both CRAN and  GitHub (https://github.com/thiyangt/tsdataleaks) and can be installed and loaded into the R session using:

\begin{verbatim}
install.packages("tsdataleaks")
library(tsdataleaks)
\end{verbatim}

or

\begin{verbatim}
devtools::install_github("thiyangt/tsdataleaks")
library(tsdataleaks)
\end{verbatim}

\subsection{Functionality}\label{Functionality}

There are three functions in the package: i) `find\_dataleaks`, ii) `viz\_dataleaks`, and iii) `reason\_dataleaks`. To demonstrate the package functions, a small time series collection with three time series is created. 

\begin{verbatim}
set.seed(2024)
x <- rnorm(15)
lst <- list(
  x = x,
  y = c(rnorm(10), x[1:5]),
  z = c(rnorm(10), x[10:15]))
\end{verbatim}

Following are the steps for detecting data leakages and visualizing the results.

\begin{figure}[htbp]
\centering
\includegraphics[width=1\textwidth]{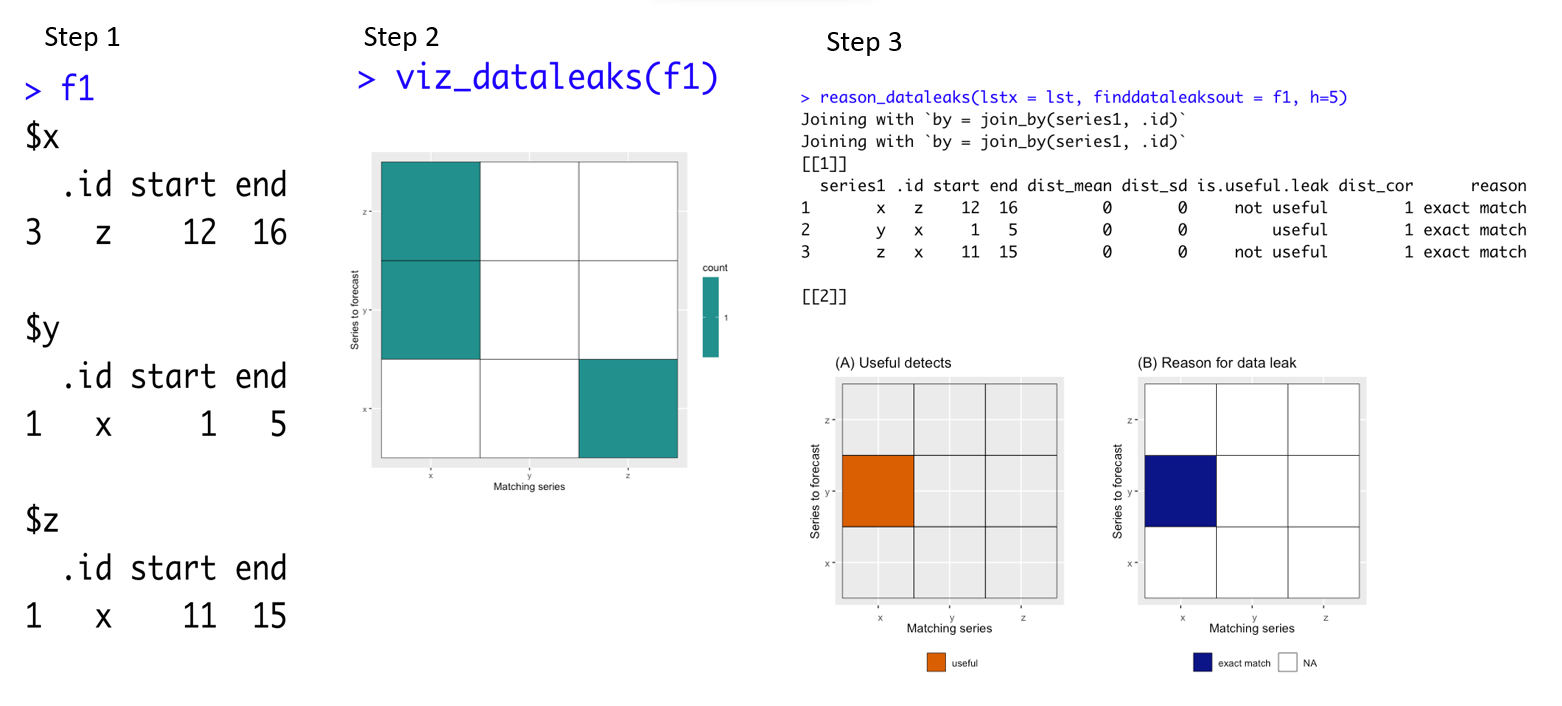}
\caption{The outputs of f1, viz\_dataleaks(f1) and reason\_dataleaks(lstx = lst, finddataleaksout = f1, h=5)}
 \label{fig:simulated}
\end{figure}

\textbf{Step 1:} The main function in the package is `find\_dataleaks`. It exploits the data leakages according to the algorithm. The inputs to the function are a list of time series collections (lstx), the length of the segment to be considered (h), and the cutoff value for the absolute value of Pearson's correlation coefficient (cutoff). The `f1` output is shown in \autoref{fig:simulated} (step1).

\begin{verbatim}
f1 <- find_dataleaks(lstx = lst, h=5, cutoff=1) 
\end{verbatim}

\textbf{Step 2:} `viz\_dataleaks` function arranges the results of `find\_dataleaks` in matrix form and visualizes them for easy understanding, as shown in \autoref{fig:simulated} (step2).

\begin{verbatim}
viz_dataleaks(f1)
\end{verbatim}

\textbf{Step 3:} `reason\_dataleaks` displays the reasons for data leaks and evaluates the usefulness of data leaks towards the winning of the competition. The inputs to the function are a list of time series collections (lstx), length of the segment to be considered (h), output of the find\_dataleaks function (finddataleaksout). The corresponding outputs are shown in \autoref{fig:simulated} (step3).

\begin{verbatim}
reason_dataleaks(lstx = lst, finddataleaksout = f1, h=5)
\end{verbatim}

According to the \autoref{fig:simulated} step 1 results, the last part of the training set in x series perfectly correlates with z series: 12–16 observations; the last part of y series correlates with x series: 1–5 observations; and the last part of z series correlates with x series: 11–15 observations. However, according to the step 3 results, only "the last part of the y series correlates with the x series: 1–5 observations" identification is useful in winning the competition. The reason is that we have x-series 6:10 observations. Hence, we can use that as the test value of the y series. "The last part of the z series correlates with x series: 11–15 observations"-This identification is not useful in winning the competition because x series: 16–20 observations are not available. In this example, all data leakages occur due to an exact match.

\hypertarget{Appication to the M1 Competition Yearly Time Series Data}{%
\section{Appication to the M1 Competition Yearly Time Series Data}\label{m1}}

\begin{figure}[htbp]
\centering
\includegraphics[width=1\textwidth]{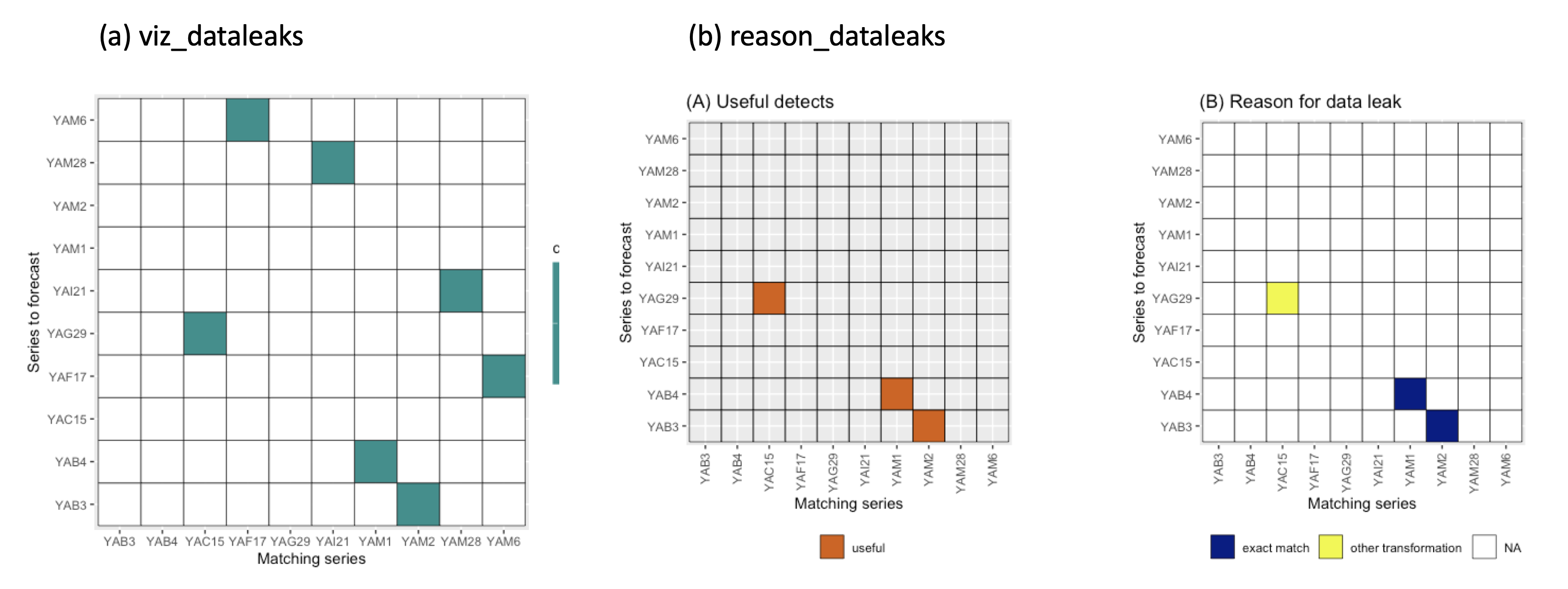}
\caption{Outputs of viz\_dataleaks(m1y\_f1) and reason\_dataleaks(M1Y\_x, m1y\_f1, h=6, ang=90)}
 \label{fig:m1y}
\end{figure}

M-competitions is a series of time-series forecasting competitions organized by Spyros Makridakis and his team \citep{makridakis2020m4}. M1-competition data is available in the package Mcomp \citealt{mcomp}. Before applying the find\_dataleaks function, all of the training sets of yearly series are stored in a list called `M1Y\_x`. In the M1 competition, the length of the test period for the yearly series is 6. Hence, the `h` value is selected as 6. The cutoff value for the absolute value of Pearson's correlation coefficient is 1. The results are shown in \autoref{fig:m1y}.

\begin{verbatim}
library(Mcomp)
data("M1")
M1Y <- subset(M1, "yearly")
M1Y_x <- lapply(M1Y, function(temp){temp$x})
m1y_f1 <- find_dataleaks(M1Y_x, h=6, cutoff = 1)
viz_dataleaks(m1y_f1)
reason_dataleaks(M1Y_x, m1y_f1, h=6, ang=90)
\end{verbatim}

According to \autoref{fig:m1y}, there are 7 data leakage detentions. Out of them, only three are useful in winning the competition. Two data leakages are due to an exact match; the other is due to a linear transformation of the form $y = mx + c$.

\hypertarget{Documentation and Examples}{%
\subsection{Documentation and Examples}\label{Documentation and Examples}}

Applications to other examples can be found in the README.md file at https://github.com/thiyangt/tsdataleaks.

\hypertarget{Conclusion}{%
\section{Conclusion}\label{Conclusion}}

The new open-source R package described in this paper enables: i) exploiting data leakages; ii) identifying the reasons for data leakage as exact matches or adding a constant or  other transformations. iii) determining whether the data leakages identified are useful in winning the forecast competition; and iv) visualizing the results. The R package tsdataleaks is a valuable tool for competition organizers to avoid data leakages, participants to detect data leakages, and the entire forecasting research community to evaluate the quality of data.

\hypertarget{Reproducibility}{%
\section{Reproducibility}\label{Reproducibility}}

Codes to generate this manuscript is available at https://github.com/thiyangt/tsdataleaks/blob/master/paper/paper.md

\bibliography{RJreferences.bib}

\address{%
Thiyanga S. Talagala\\
Department of Statistics\\%
Faculty of Applied Sciences\\ University of Sri Jayewardenepura, Sri
Lanka\\
\url{https://github.com/thiyangt/ceylon}\\%
\textit{ORCiD: \href{https://orcid.org/0000-0002-0656-9789X}{0000-0002-0656-9789X}}\\%
\href{mailto:ttalagala@sjp.ac.lk}{\nolinkurl{ttalagala@sjp.ac.lk}}%
}

\end{article}

\end{document}